\documentclass[epj]{svjour}

\usepackage{graphicx}
\usepackage{fancyhdr}
\def\makeheadbox{{
 }}
\def\mytitle{My title} 
\def\myauthors{My name}  
\def\mytype{My type of session}
\def\mysession{My session}

\def\mytitle{Search for Higgs Bosons in H to WW Decays at the Tevatron} 
\def\myauthors{Nils Krumnack, Baylor University}    
\def\mytype{Contributed Talk}    
\def\mysession{Colliders - Higgs Phenomenology}

\begin{document}
\title{\mytitle}
\subtitle{}
\author{Nils Krumnack\inst{1} on behalf of the CDF and D\O{} Collaborations
}                     
%
%
\institute{Baylor University
}
%
\date{}
\abstract{
We present the results of searches by the CDF and D\O{} Collaborations for Higgs boson production in $p\bar{p}$ collisions at $\sqrt{s}=1.96\,{\rm TeV}$.  The searches are performed in the $WW^*$ channel with $1\,{\rm fb}^{-1}$ of data.  In the absence of signal the results are used to set a limit on the Higgs production cross section times branching ratio.
} 
\maketitle
%

\section{Introduction}

The Higgs boson, the last undiscovered particle of the standard model,
explains the origin of mass and electro-weak symmetry breaking.
Within the standard model the only free parameter of the Higgs boson
is its mass $m_H$.  The combination of precision measurements
indicates that its mass should be less than $144\,{\rm GeV}$
\cite{LEP2}.  Previous direct searches at LEP have set a lower limit
on the Higgs mass at $114\,{\rm GeV}$ \cite{LEP1}.  Here we present
the result of searches in the $WW^*\rightarrow l\nu l\nu$ decay
channel, which covers the whole range of allowed Higgs masses and is
particularly sensitive in the high mass range.

The main signature of the $WW^*\rightarrow l\nu l\nu$ decay channel is
two leptons of opposite charge and missing transverse energy
caused by the neutrinos escaping the detector undetected.

\section{The D\O{} measurement}

\begin{figure}
  \caption{The $\Delta\phi$ spectrum after applying all cuts for D\O{}.  The limit is extracted from the region $\Delta\phi<2$.}
  \begin{center}
    \includegraphics[width=.48\linewidth]{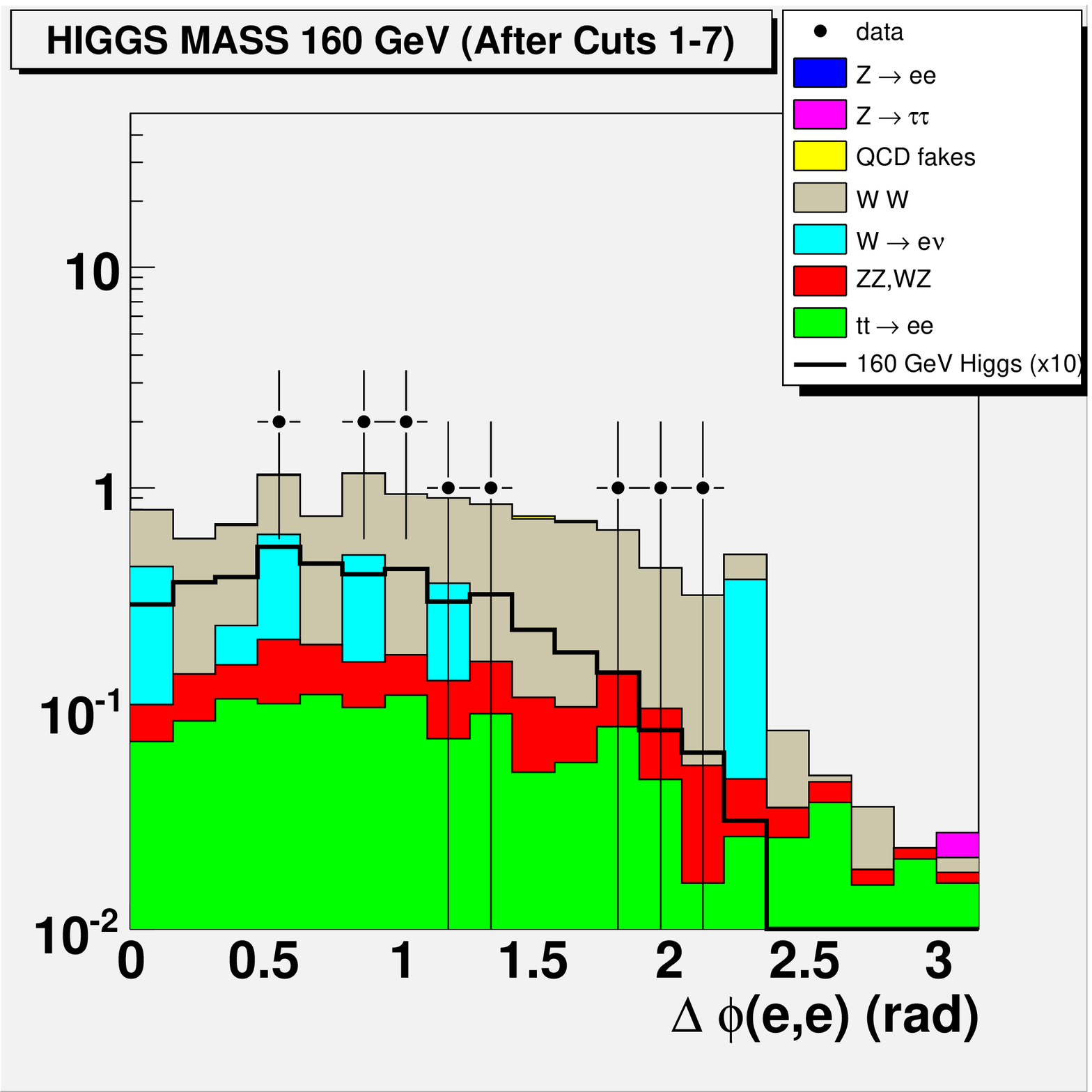}
    \includegraphics[width=.48\linewidth]{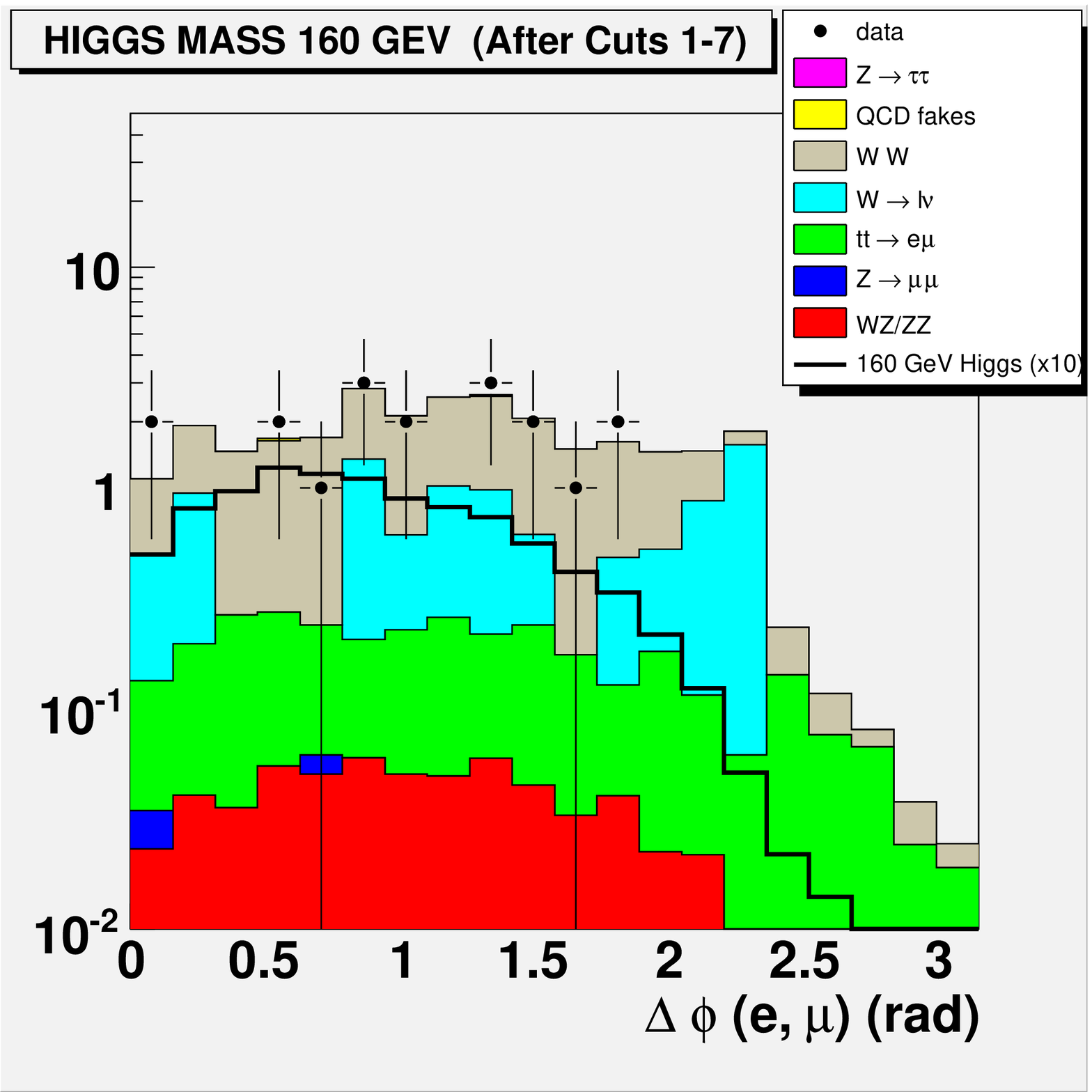}
    \includegraphics[width=.48\linewidth]{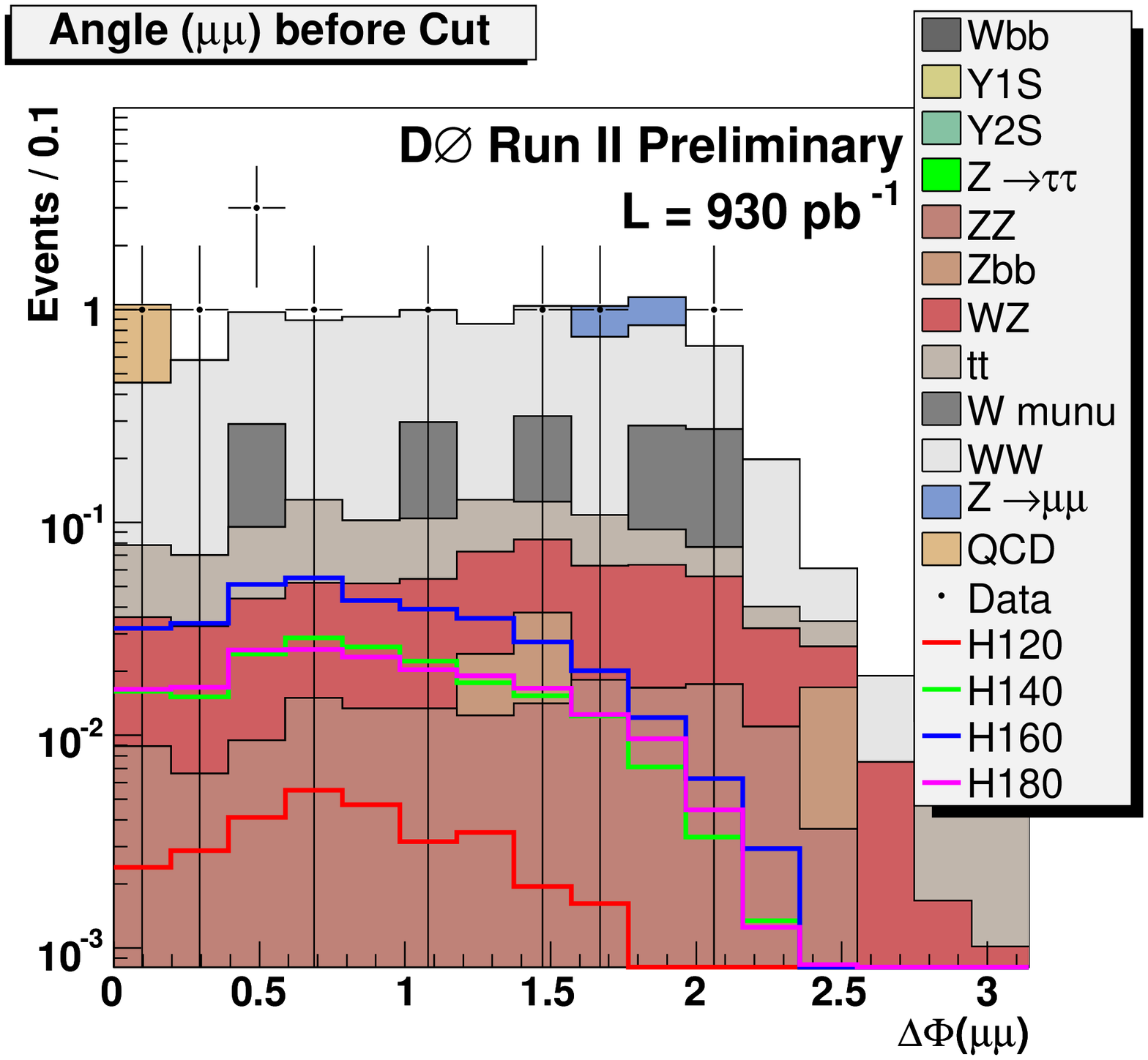}
  \end{center}
  \label{D0 control 3}
\end{figure}

\begin{figure}
  \caption{The expected and observed D\O{} limit on the cross section
    as a function of mass.  For reference the standard model cross
    section and the cross section for a fourth generation model are
    shown as well.  The measurement already excludes the fourth
    generation model for some Higgs masses and is within a factor of 4
    from the standard model prediction at 160~GeV.}
  \begin{center}
    \includegraphics[width=\linewidth]{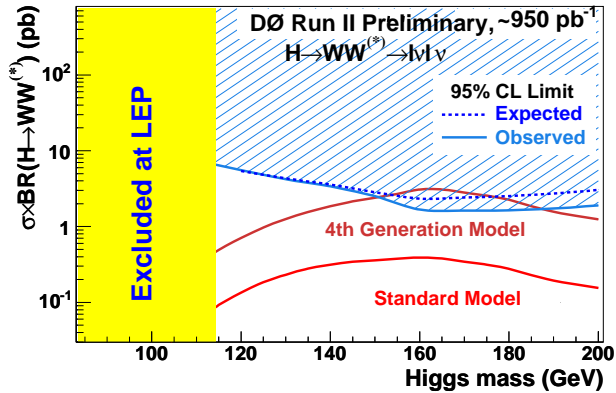}
  \end{center}
  \label{D0 result}
\end{figure}

The D\O{} measurement uses a simple cut-based
analysis~\cite{D0A}\cite{D0B}.  The cuts are listed in Table~\ref{D0
  cuts}.  Control plots for the data sample are in Figure~\ref{D0
  control 1}.  To avoid the uncertainties associated with the
luminosity measurement, the background is normalized to the $Z$ mass
peak which is calculated at NNLO and fitted to the observed data.  The
final selection variable is the $\Delta\phi$ between the two leptons.
The distribution of that variable after applying all selection cuts
can be seen in Figure~\ref{D0 control 3}.  No excess over background
is observed and the data are used to set a limit on the Higgs
production cross section (Figure~\ref{D0 result}).

\section{The CDF measurement}

CDF has performed two measurements in this channel, both of them using
multivariate techniques.  The first one uses a neural network
approach~\cite{CDFNN} and the second one uses a matrix element
technique~\cite{CDFME}.  These techniques are complementary and work
is being performed to integrate them into a combined limit.  By
construction, the matrix element method only takes leading order
effects into account, while the neural network approach can also take
into account next-to-leading-order effects as emulated by the parton
shower models.

\begin{figure}
  \caption{The expected and observed cross section limit for the CDF
    neural network analysis.  The limit is normalized to the standard
    model prediction.  The prediction for a fourth generation model is
    shown as well.  The measurement already excludes the fourth
    generation model for some Higgs masses and is within a factor of 5
    from the standard model prediction at 160~GeV.}
  \begin{center}
    \includegraphics[width=\linewidth]{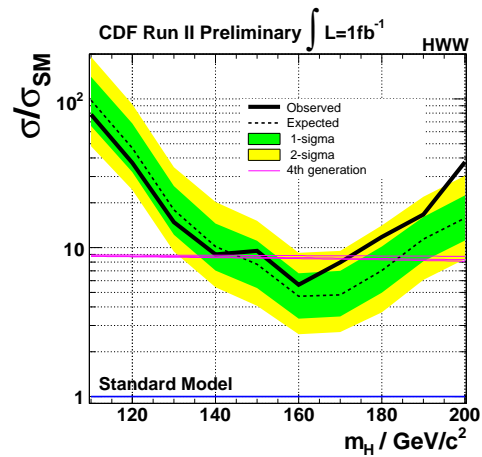}
  \end{center}
  \label{CDF NN result}
\end{figure}

The neural network analysis uses the cuts in Table~\ref{CDF NN cuts}.
Control plots for the event selection can be seen in Figure~\ref{CDF
  NN control 1}.  For the events passing the cuts, a neural network is
trained with the input variables in Table~\ref{CDF NN var} to
discriminate between signal and background.  The resulting event
discriminant can be seen in Figure~\ref{CDF NN control 2}.  This
procedure is repeated for every mass point and for each mass point a
limit is calculated (Figure~\ref{CDF NN result}).

\begin{figure}
  \caption{The event discriminant for the CDF ME analysis in the low
    (top) and high (bottom) signal-to-background region.  The inlay
    shows a magnification of the events with high discriminant values.
    For comparison, the standard model Higgs prediction is shown
    (scaled up by a factor of 10).  Each bin is treated as a separate
    counting experiment for the extraction of the limit.}
  \begin{center}
    \includegraphics[width=\linewidth]{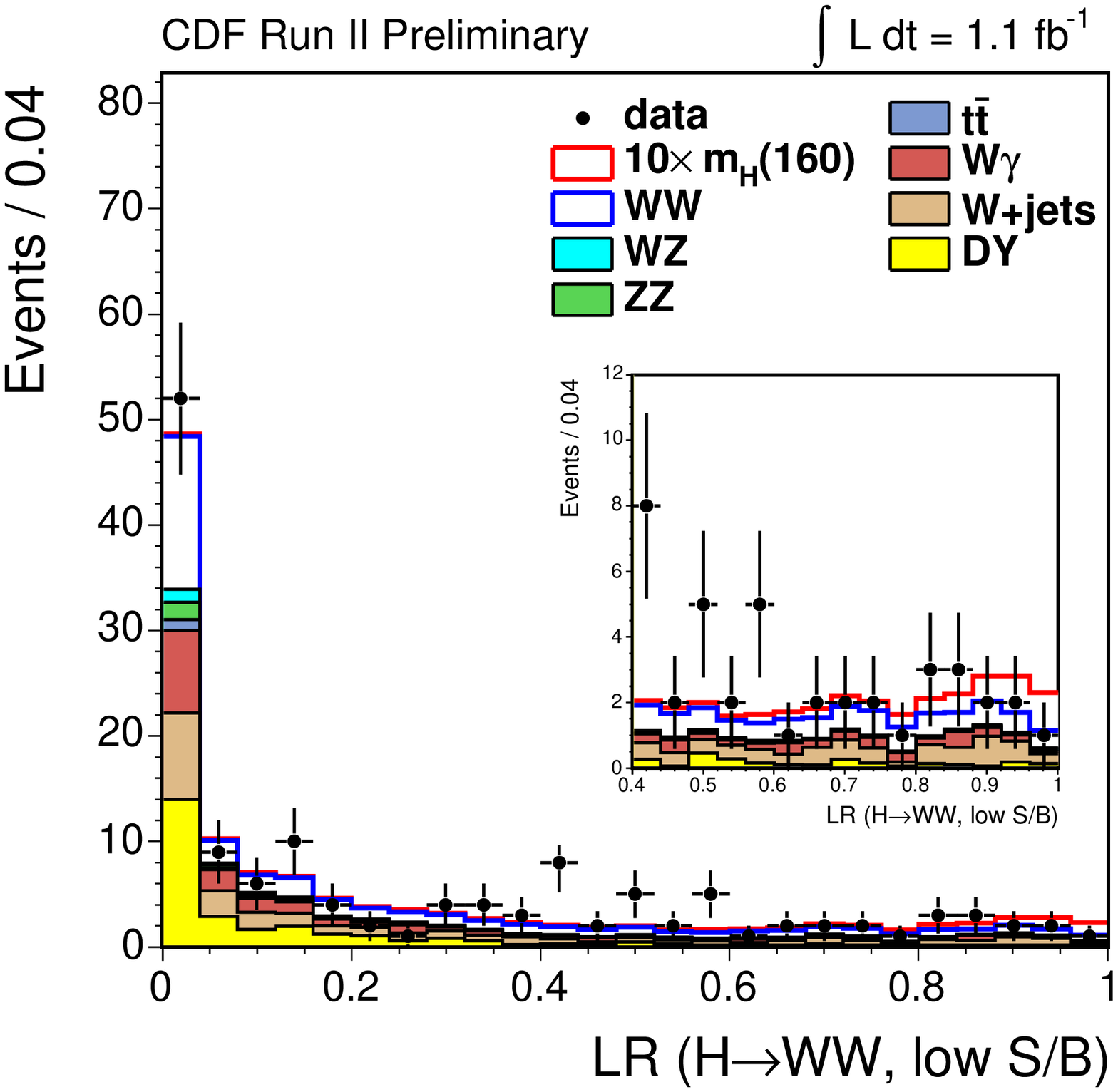}
    \includegraphics[width=\linewidth]{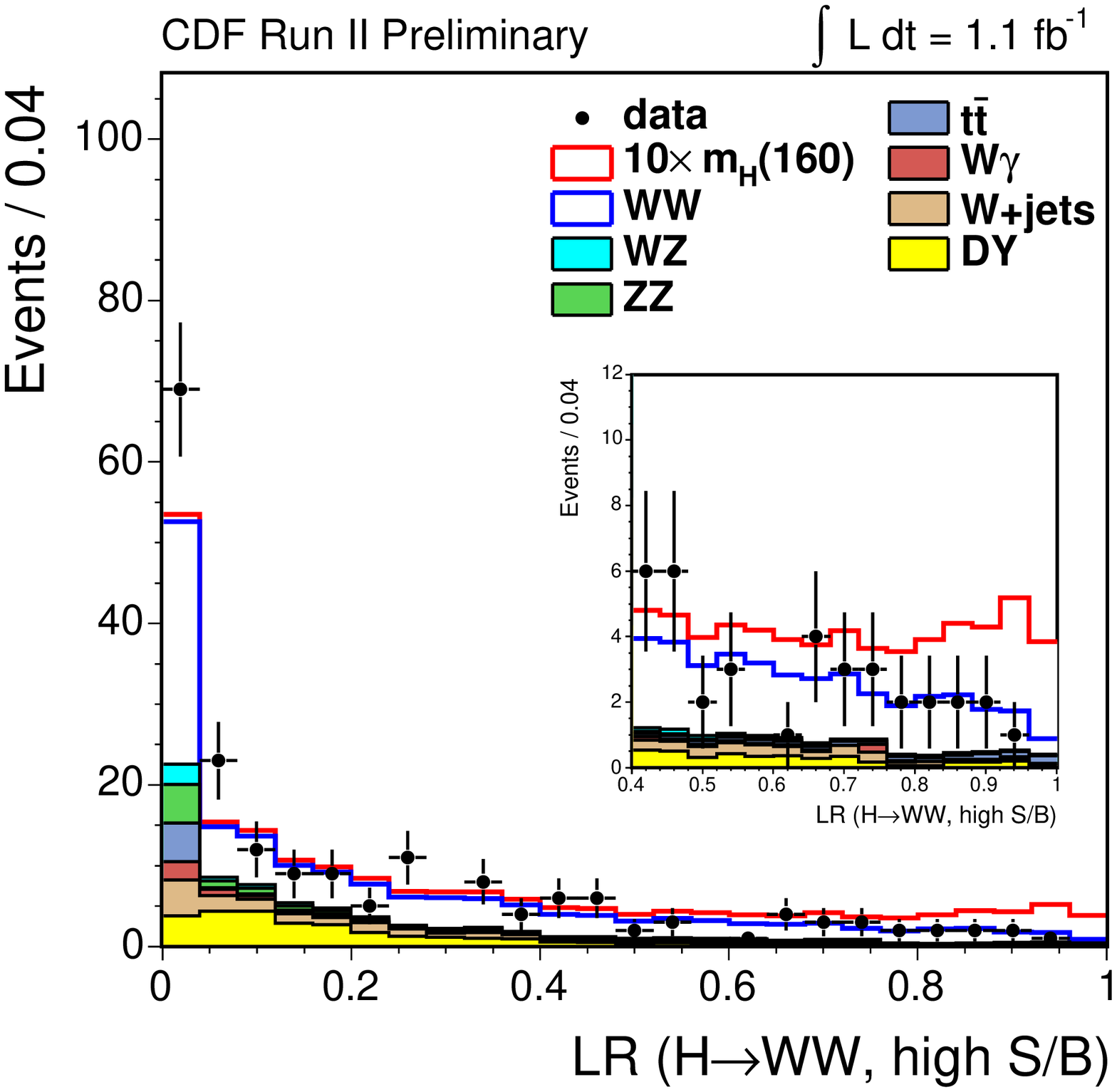}
  \end{center}
  \label{CDF ME control 1}
\end{figure}

\begin{figure}
  \caption{The expected and observed cross section limit for the CDF
    matrix element analysis.  The limit is normalized to the standard
    model prediction.  The measurement is within a factor of 4 from
    the standard model prediction at 160~GeV.}
  \vspace{1mm}
  \begin{center}
    \includegraphics[width=\linewidth]{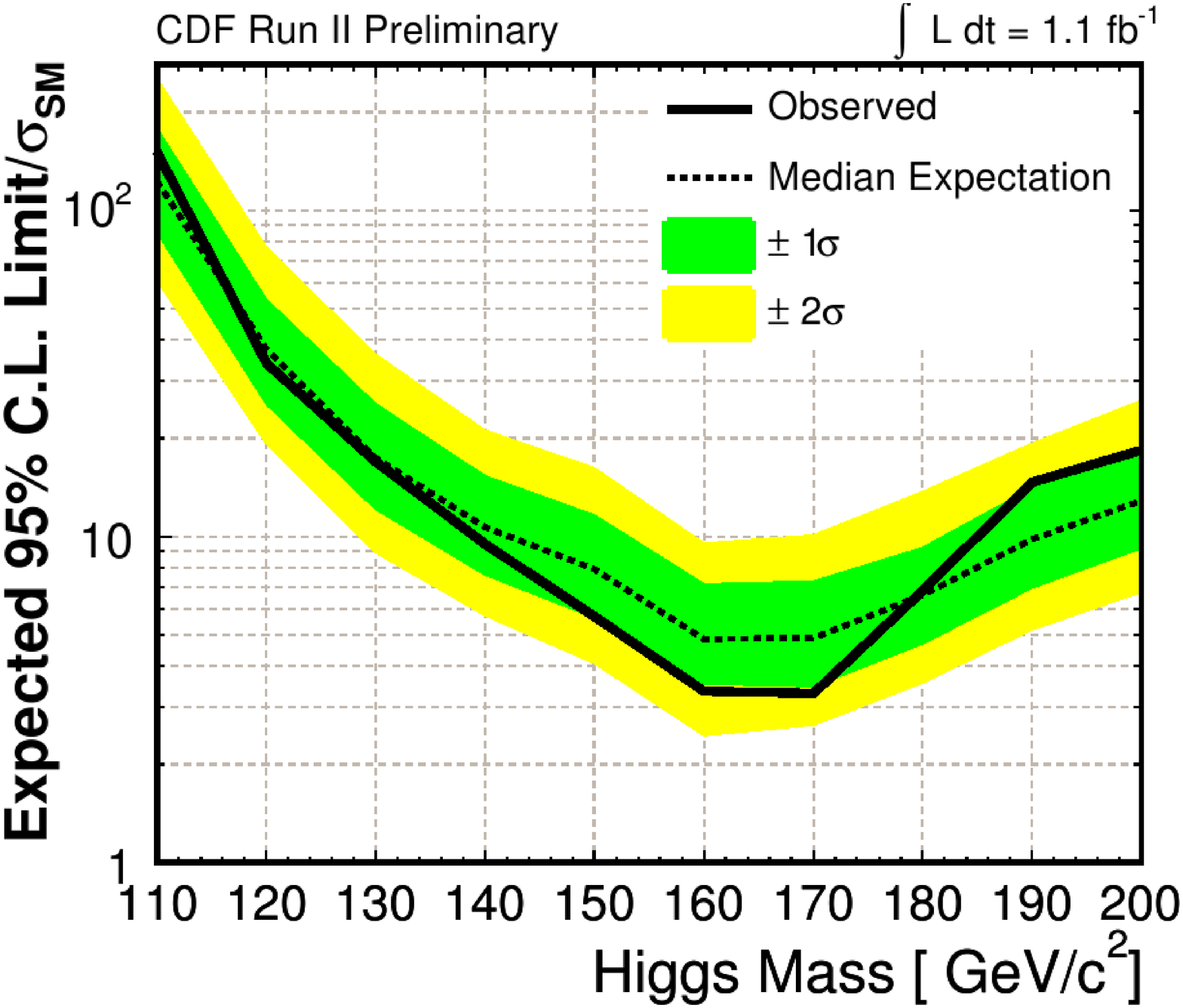}
  \end{center}
  \label{CDF ME result}
\end{figure}

The basic idea of the matrix element technique is to use LO matrix
elements to calculate event probabilities.  For each event and process
the LO matrix element is integrated over phase space:
\begin{equation}
P_m(\vec x)=\int\frac{d\sigma_m(\vec y)}{d\vec y}\epsilon(\vec
y)G(\vec x,\vec y)d\vec y
\end{equation}
where $\frac{d\sigma_m(\vec y)}{d\vec y}$ is the matrix element,
$\epsilon(\vec y)$ is the efficiency and $G(\vec x,\vec y)$ is the
resolution.  The matrix element analysis uses the cuts in
Table~\ref{CDF ME cuts}.  The data are divided into a high
signal-to-background and a low signal-to-background region (see
Figure~\ref{CDF ME control 1}).  For each mass point a separate event
discriminant is calculated and a separate limit is calculated (see
Figure~\ref{CDF ME result}).

\section{Combining Measurements}

\begin{figure}
  \caption{The combined limits for both experiments and the Tevatron
    as a whole.  The CDF combination uses an older version of the
    $H\rightarrow WW$ analysis, which only used $300\,{\rm pb}^{-1}$.
    Except for the low mass region the limit is dominated by the
    $H\rightarrow WW$ channel and has its maximum sensitivity at
    160~GeV.}
  \begin{center}
    \includegraphics[width=\linewidth]{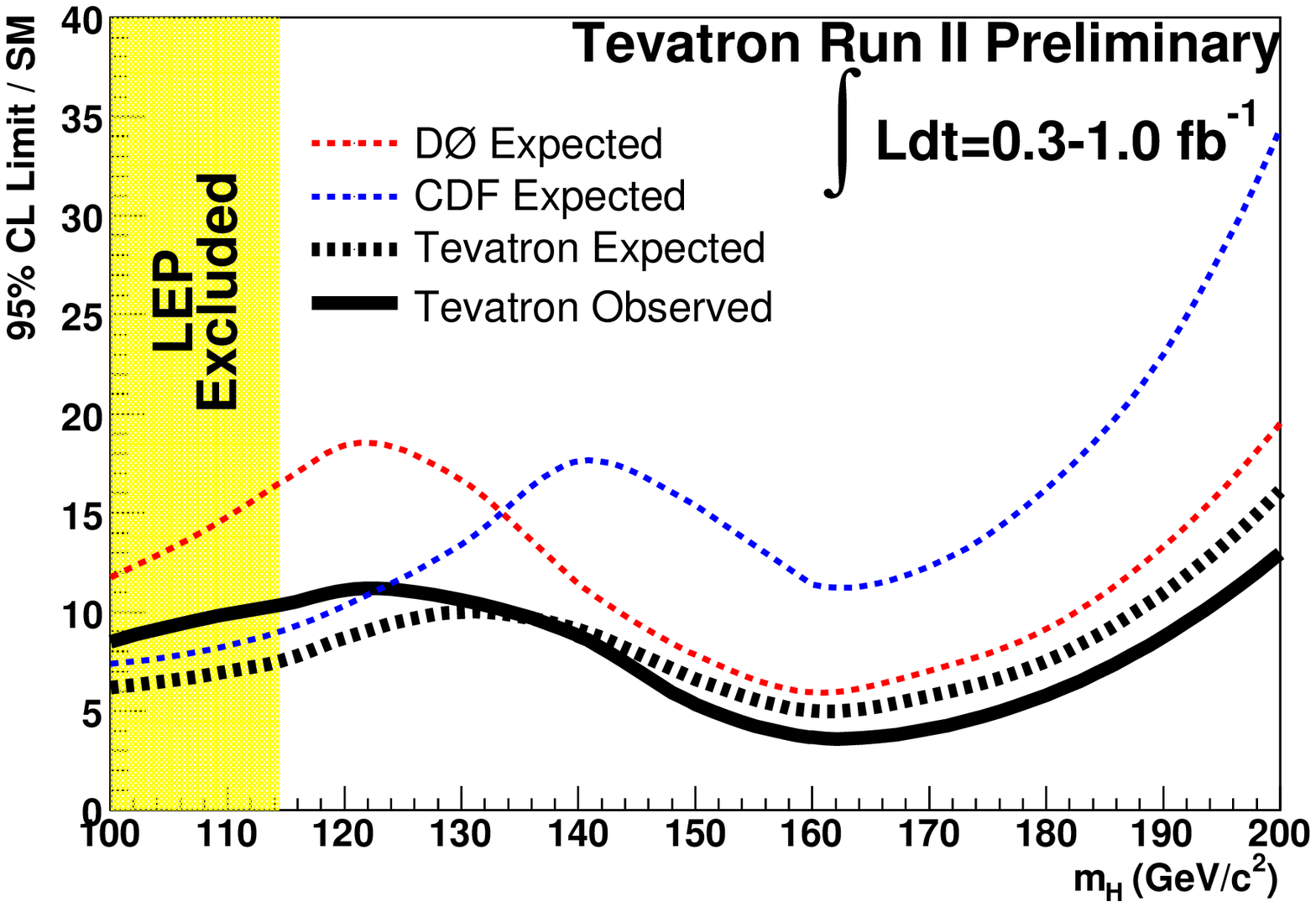}
  \end{center}
  \label{combination}
\end{figure}

For each experiment the obtained limits are combined with the limits
from other search channels.  In addition, the limits from CDF and
D\O{} are also combined to yield an overall Tevatron limit (see
Figure~\ref{combination}).

\begin{table*}
\caption{Cut values used in the D\O{} analysis.  $p_{T,i}$ is the
  transverse momentum of lepton $i$ (ordered by decreasing $p_T$).
  $m_ll$ is the invariant mass of the two-lepton system.
  $\Delta\phi_{ll}$ is the difference in azimuthal angle between the
  two leptons.  $m_{T,\min}(l,\not\!\!{E_T})$ is the minimum
  transverse mass of the missing transverse momentum and one of the
  leptons.}
\begin{center}
\begin{tabular}{|c|c|c|c|}
\hline
&${\rm ee}$&${\rm e}\mu$&$\mu\mu$\cr\hline\hline
lepton ID&\multicolumn{3}{c|}{$p_{T,1}>15, p_{T,2}>10, m_{ll} > 15, {\rm isolation}$}\cr\hline
$\not\!\!{E_T}$&\multicolumn{3}{c|}{$\not\!\!{E_T}>20, {\rm significance}(\not\!\!{E_T})>7$}\cr\hline
$m_{ll}<x$&$\min(m_H/2,80)$&$m_H/2$&$80$\cr\hline
$p_{T,1}+p_{T,2}+\not\!\!{E_T}$&\multicolumn{2}{c|}{$m_H/2+20<x<m_H$}&$100<x<160$\cr\hline
$m_{T,\min}(l,\not\!\!{E_T})$&\multicolumn{2}{c|}{$x>15+m_H/4$}&$x>55$\cr\hline
$H_T=\sum p_T^{\rm jet}$&\multicolumn{2}{c|}{$H_T<100$}&$H_T<70$\cr\hline
$\Delta\phi_{ll}$&\multicolumn{3}{c|}{$\Delta\phi_{ll}<2.0$}\cr\hline
\end{tabular}
\end{center}
\label{D0 cuts}
\end{table*}%

\begin{table*}
\caption{Cut values used in the CDF neural net analysis (left) and the
  input variables for the neural net (right).}
\begin{center}
\begin{tabular}{|c|}\hline
$p_{T,1}>20,p_{T,2}>10$\cr\hline
lepton isolation\cr\hline
$m_{ll}>16$\cr\hline
$n_{\rm jet}=0$ or $n_{\rm jet}=1,E_T^{\rm jet}<55$ or $n_{\rm jet}=2, E_T^{\rm jet}<40$\cr\hline
opposite charge leptons\cr\hline
neural net for Drell-Yan suppression\cr\hline
\end{tabular}\quad
\begin{tabular}{|c|c|}
\hline
$p_{T,1}$&$p_{T,1}+p_{T,2}+\not\!\!E_T$\cr\hline
$p_{T,2}$&$m_{ll}$\cr\hline
$n_{\rm jets}$&$\Delta\phi_{\rm min}(\not\!\!E_T,{\rm lepton\ or\ jet})$\cr\hline
$E_{T,1}^{\rm jet}$&$\Delta\phi_{ll}$\cr\hline
$E_{T,2}^{\rm jet}$&$\sqrt{\Delta\eta_{ll}^2+\Delta\phi_{ll}^2}$\cr\hline
$\not\!\!E_T$&$\not\!\!E_T/(p_{T,1}+p_{T,2}+\not\!\!E_T)$\cr\hline
\end{tabular}
\end{center}
\label{CDF NN var}
\label{CDF NN cuts}
\end{table*}%

\begin{table}
\caption{Cut values used in the CDF matrix element analysis.}
\begin{center}
\begin{tabular}{|c|}\hline
$p_{T,1}>20, p_{T,2}>10$\cr\hline
$25<\not\!\!{E_{T,rel}}=\not\!\!{E_T}\cdot\sin(\min(\pi/2,\Delta\phi(\not\!\!{E_T},{\rm lepton\ or\ jet}))$\cr\hline
$\not\!\!E_T\sqrt{\sum E_T}>2.5$\cr\hline
$n_{\rm jets}<2$\cr\hline
$m_{ll}>25$\cr\hline
trilepton veto\cr\hline
\end{tabular}
\end{center}
\label{CDF ME cuts}
\end{table}%

\begin{figure}
  \caption{The invariant mass spectrum in the preselection sample for
    D\O{}.  To avoid the uncertainties in the luminosity measurement,
    the background prediction is normalized to the $Z$ mass peak.}
  \begin{center}
    \includegraphics[width=.48\linewidth]{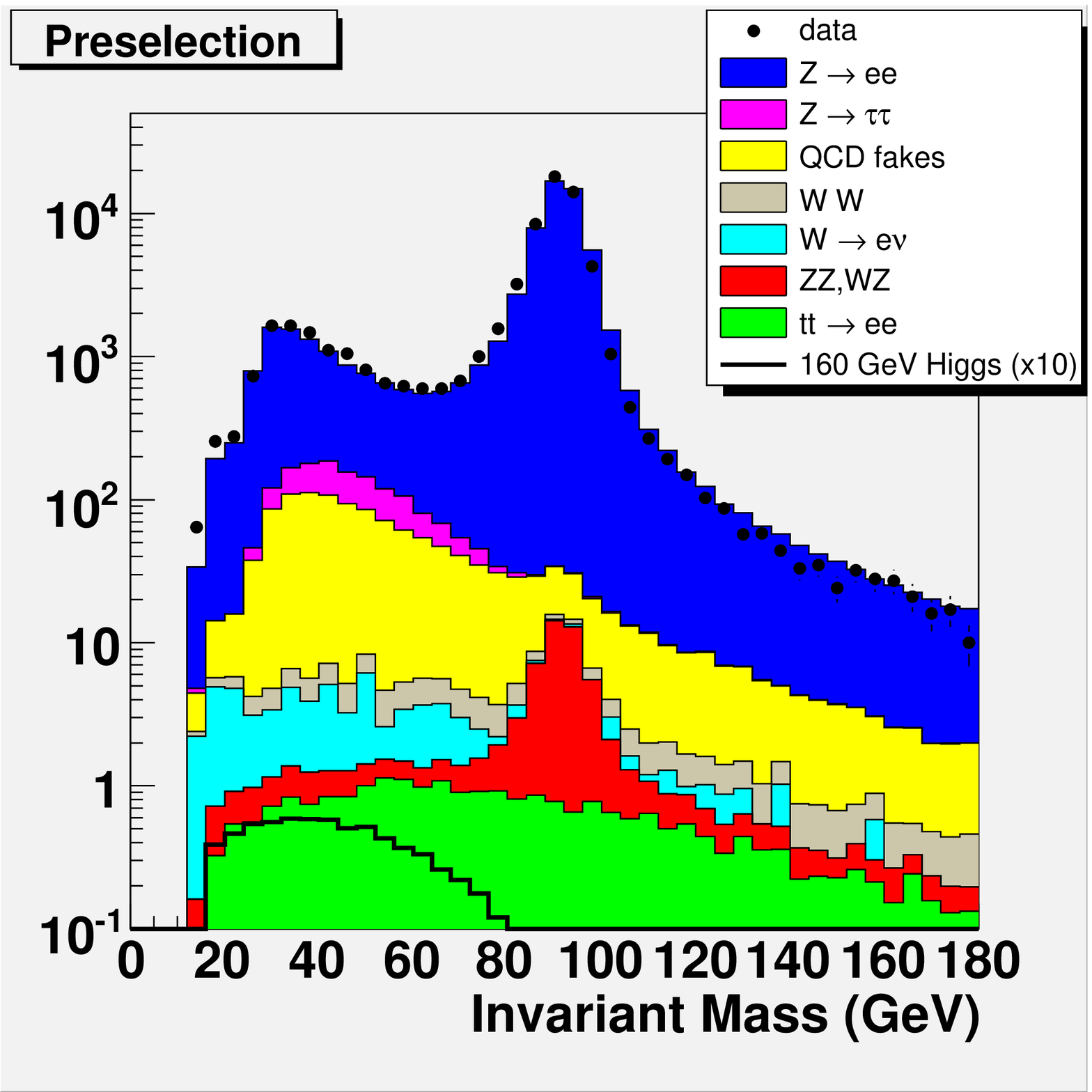}
    \includegraphics[width=.48\linewidth]{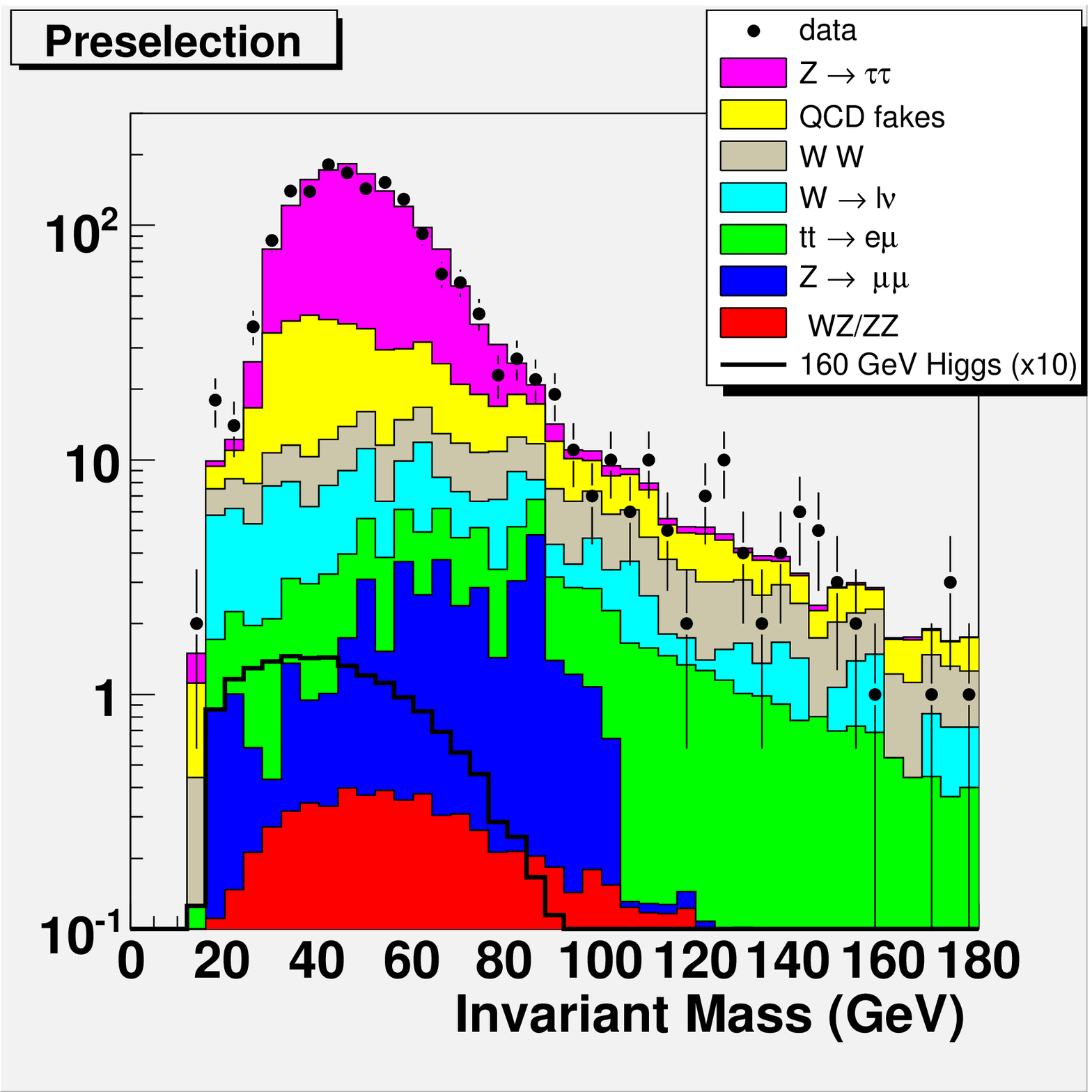}
    \includegraphics[width=.48\linewidth]{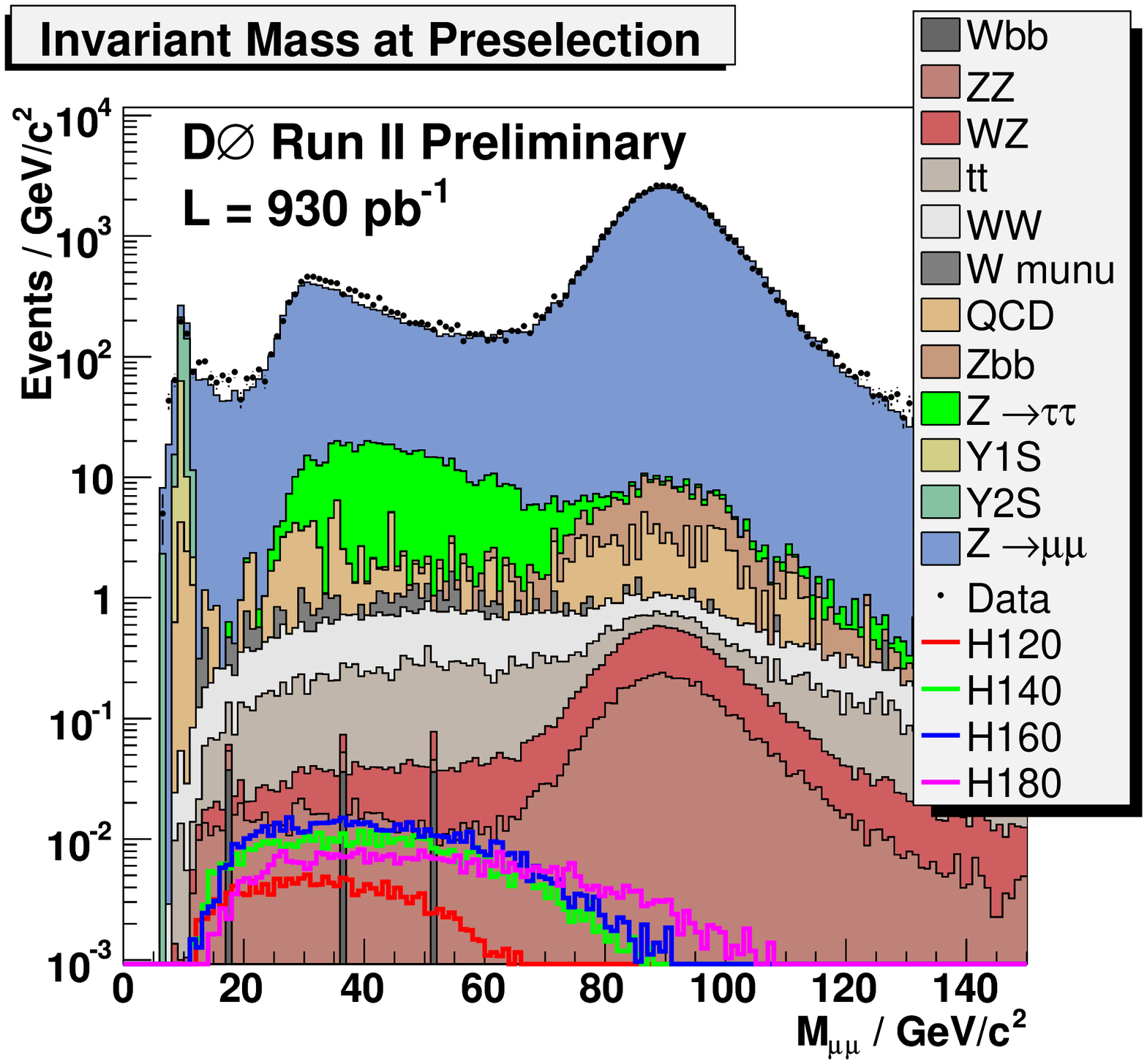}
  \end{center}
  \label{D0 control 1}
\end{figure}

\begin{figure}
  \caption{Control plots for the CDF neural network analysis after
    applying the selection cuts.  The background model describes the
    signal well and the variables allow discrimination between signal
    and background.}
  \begin{center}
    \includegraphics[width=.48\linewidth]{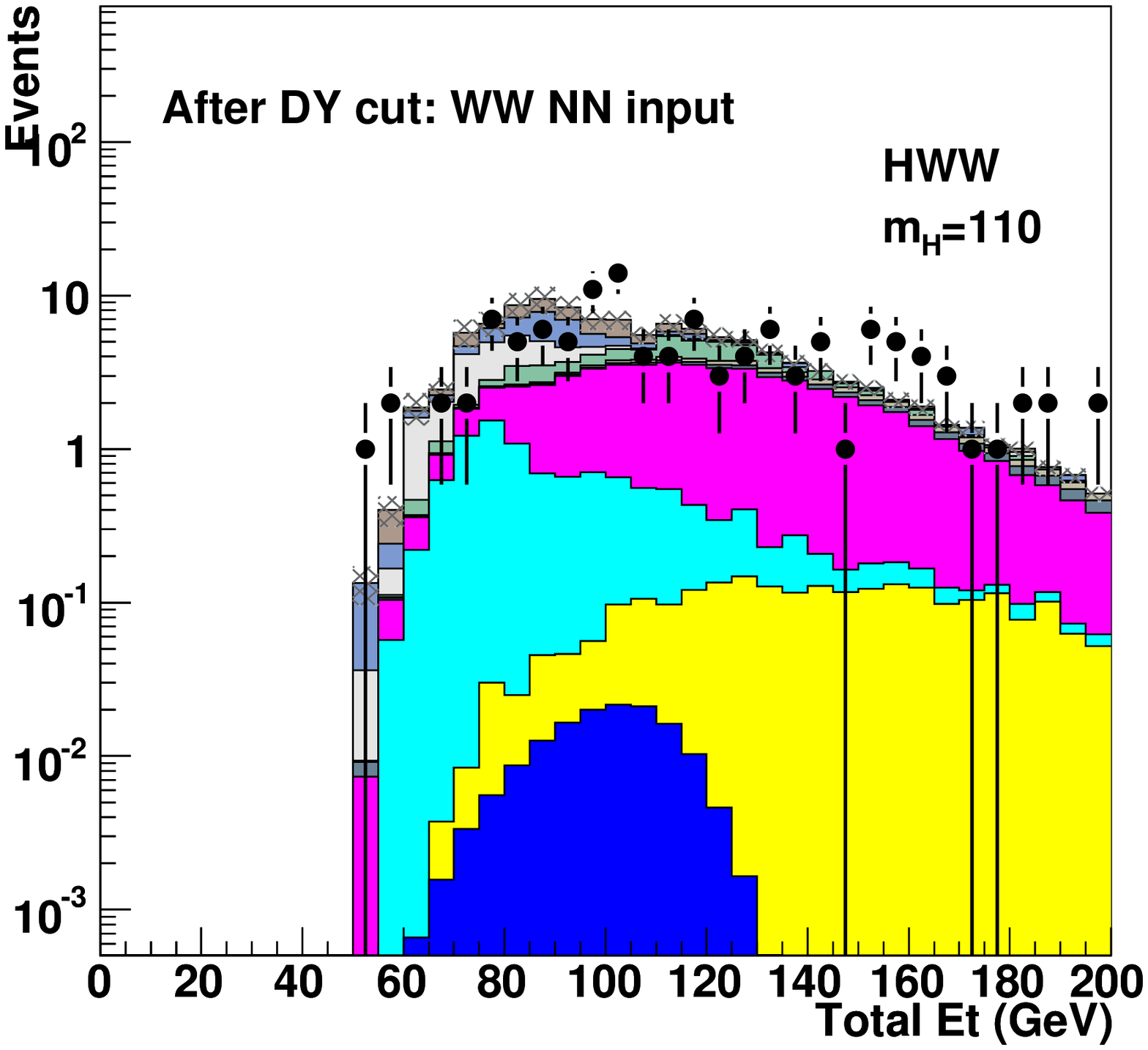}
    \includegraphics[width=.48\linewidth]{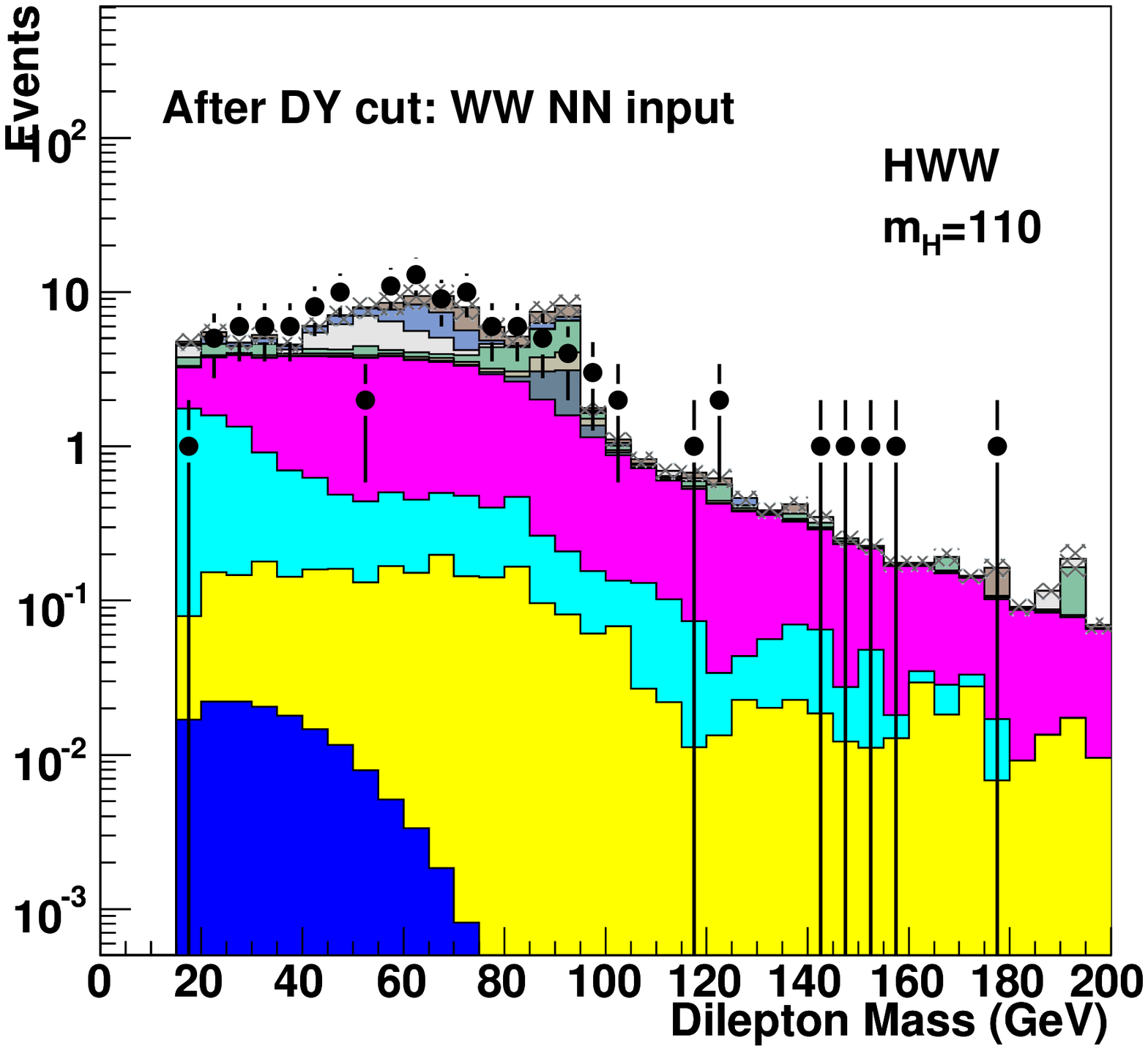}
    \includegraphics[width=.48\linewidth]{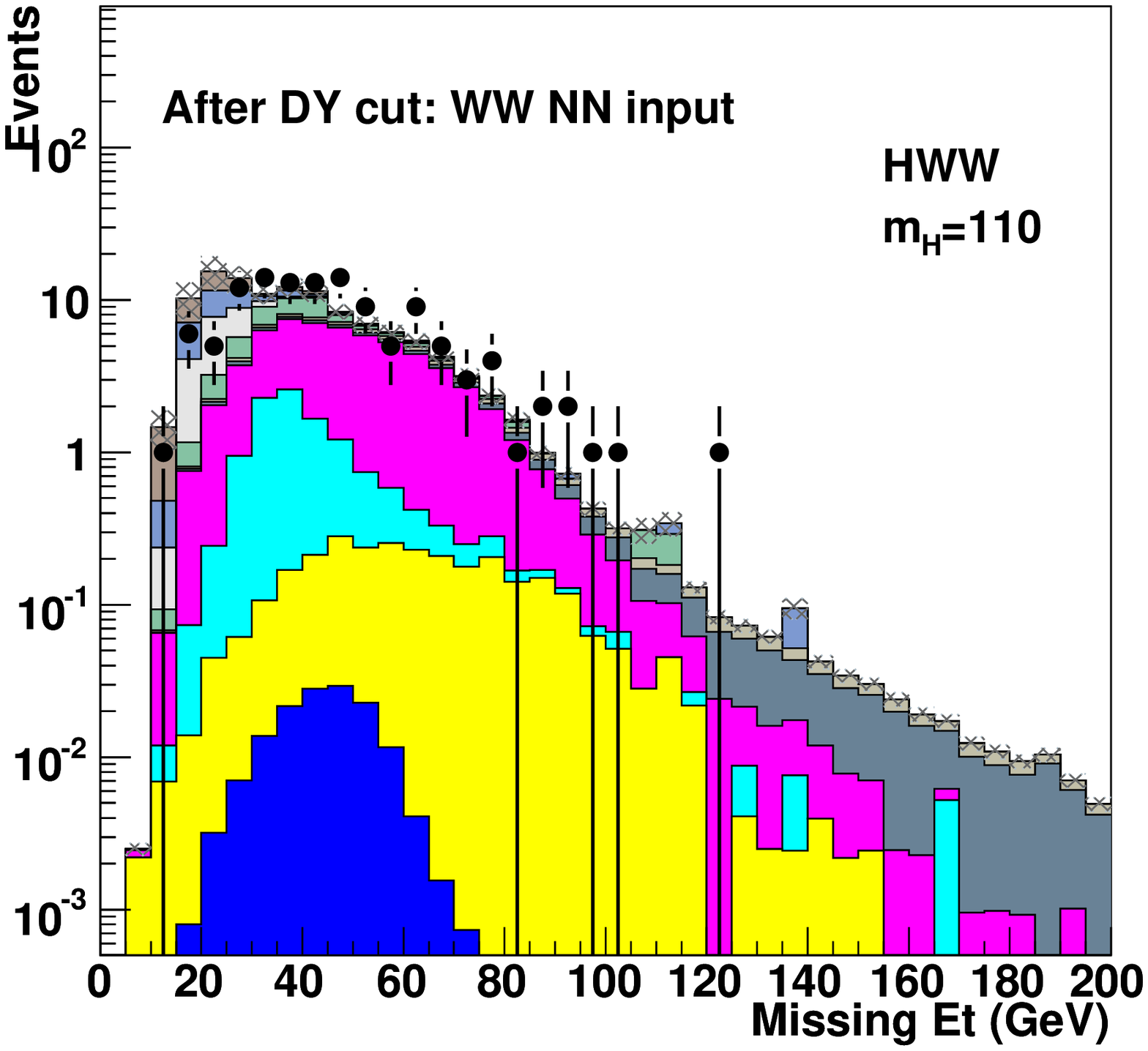}
    \includegraphics[width=.5\linewidth]{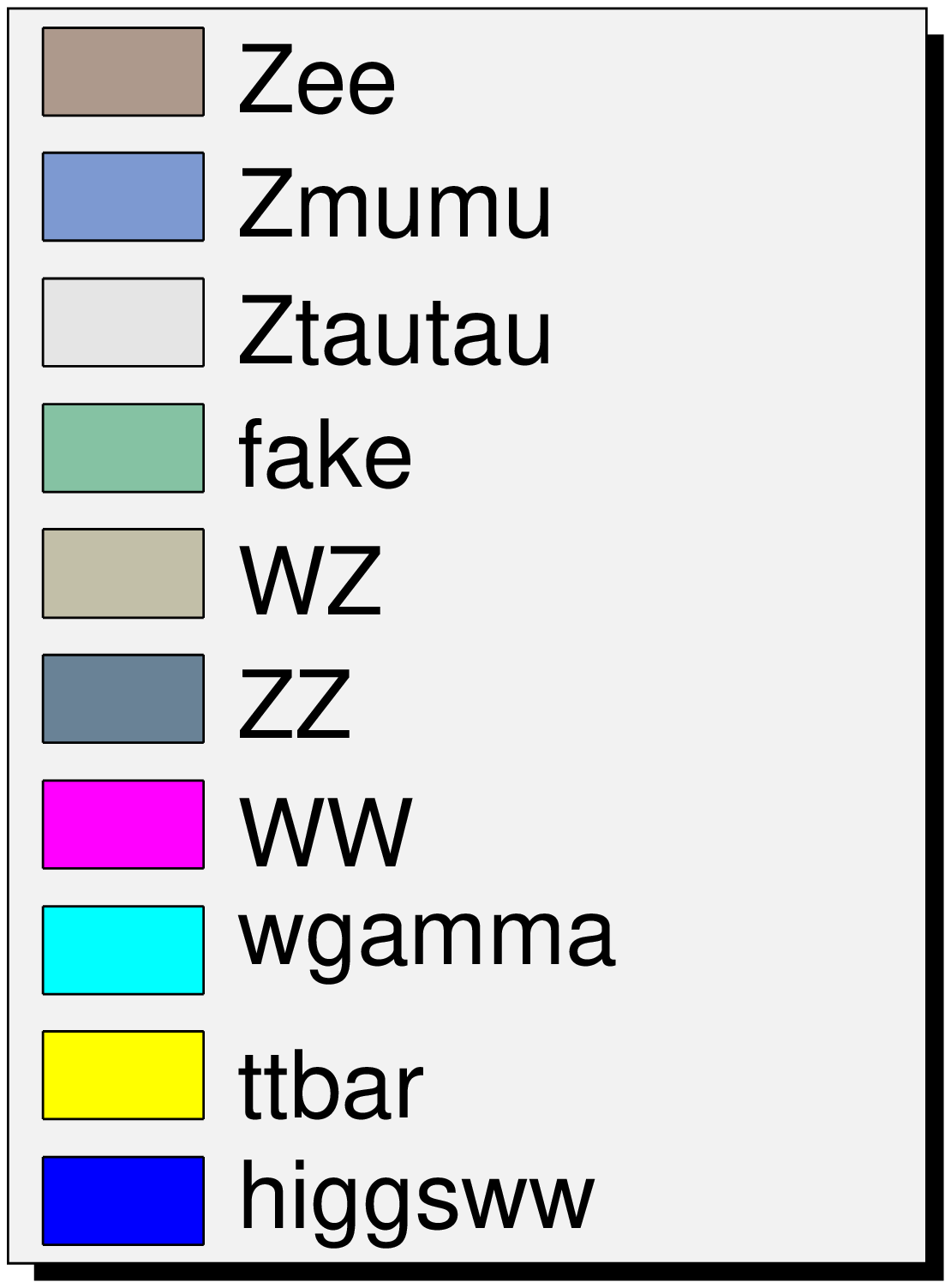}
  \end{center}
  \label{CDF NN control 1}
\end{figure}

\begin{figure}
  \caption{The event discriminant (i.e. network output) for the CDF neural network analysis.  The shaded area indicates the uncertainty on the background prediction.  Each bin is treated as a separate counting experiment to determine the limit.  A legend to the plots can be found in Figure~\ref{CDF NN control 1}.}
  \begin{center}
    \includegraphics[width=\linewidth]{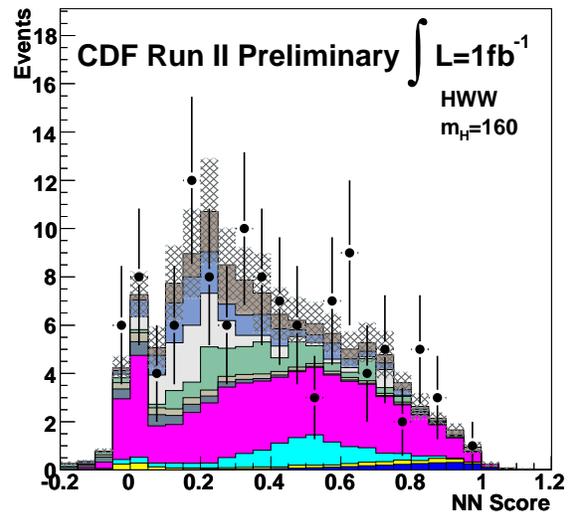}
  \end{center}
  \label{CDF NN control 2}
\end{figure}

%
%

\end{document}